\documentclass[aps,twocolumn,showpacs,preprintnumbers,nofootinbib,floats]{revtex4}\def\@cite#1#2{\textsuperscript{[{#1\if@tempswa , #2\fi}]}}
\usepackage{mathrsfs}
\usepackage{longtable,lscape}
\usepackage{txfonts}
\usepackage{amssymb}
\usepackage{indentfirst}
\usepackage{graphicx,booktabs}
\usepackage{float}
\usepackage{setspace}
\usepackage{epstopdf}

\usepackage{multirow}
\usepackage{color}
\usepackage{epsfig}

\begin{document}


\title{The hidden-charm strong decays of the $Z_c$ states}

\author{Li-Ye Xiao$^{1}$~\footnote {E-mail: lyxiao@ustb.edu.cn}, Guang-Juan Wang$^{2,3}$~\footnote {E-mail: wgj@pku.edu.cn}, and Shi-Lin Zhu$^{2,3,4}$~\footnote {E-mail: zhusl@pku.edu.cn}}

\affiliation{ 1)School of Mathematics and Physics, University of Science and Technology Beijing,
Beijing 100083, China } \affiliation{ 2) School of Physics and State
Key Laboratory of Nuclear Physics and Technology, Peking University,
Beijing 100871, China} \affiliation{ 3) Center of High Energy
Physics, Peking University, Beijing 100871, China}  \affiliation{ 4)
Collaborative Innovation Center of Quantum Matter, Beijing 100871,
China}


\begin{abstract}
Inspired by BESIII's measurement of the decay
$Z_c(3900)^{\pm}\rightarrow \rho^{\pm}\eta_c$, we calculate the
branching fraction ratio between the $\rho\eta_c$ and $\pi J/\psi$
decay modes for the charged states $Z_c(3900)$, $Z_c(4020)$ and
$Z_c(4430)$ using a quark interchange model. Our results show that
(i) the ratio
$R_{Z_c(3900)}=\frac{\mathcal{B}(Z_c(3900)^{\pm}\rightarrow
\rho^{\pm}\eta_c)}{\mathcal{B}(Z_c(3900)^{\pm}\rightarrow
\pi^{\pm}J/\psi)}$ is 1.3 and 1.6 in the molecular and tetraquark
scenarios respectively, which is roughly consistent with the
experimental data $R^{\mathrm{exp}}=2.2\pm0.9$. (ii) The ratios
$\frac{\Gamma[Z_c(3900)\rightarrow
\rho\eta_c]}{\Gamma[Z_c(4020)\rightarrow \rho\eta_c]}$ and
$\frac{\Gamma[Z_c(3900)\rightarrow \pi
J/\psi]}{\Gamma[Z_c(4020)\rightarrow \pi J/\psi]}$ are about 12.5
and 24.2, respectively in the molecular scenario. In contrast, these
ratios are about 1.2 in the tetraquark scenario. The non-observation
of the $Z_c(4020)$ signal in the $\pi J/\psi$ decay mode strongly
indicates that $Z_c(3900)$ and $Z_c(4020)$ are molecule-like signals
which arise from the $D^{(*)}{\bar D}^{(*)}$ hadronic interactions.
\end{abstract}


\maketitle

\section{Introduction}

Since 2003 the Belle Collaboration observed the first
charmonium-like state~\cite{Choi:2003ue}, $X(3872)$, an explosion in
the observation of new hadronic states began. Dozens of
charmonium-like states (or XYZ states)~\cite{Tanabashi:2018oca} have
been reported by several major experimental collaborations such as
BESIII, LHCb, Belle, BaBar, CDF and so on; see
Ref.~\cite{Chen:2016qju,Liu:2019zoy,Brambilla:2019esw,Guo:2017jvc,Esposito:2016noz,Hosaka:2016pey}
for a review. The properties of the charged $Z_c$ states cannot be
explained by the naive quark model and make them manifestly exotic.
It seems that we have a new zoo of exotic hadrons. How to understand
their internal structures remains a great challenge.

The charged charmonium-like state
$Z_c(4430)$~\cite{Choi:2007wga,Aubert:2008aa,Mizuk:2009da,Chilikin:2013tch,Chilikin:2014bkk,Aaij:2014jqa}
was first observed $Z_c$ in 2007, which has trigged extensive
theoretical speculations, such as the molecular
state~\cite{Ma:2014zua}, the first radial tetraquark
excitation~\cite{Maiani:2007wz,Ebert:2008kb,Maiani:2014aja},
threshold cusp effects~\cite{Bugg:2008wu}, and triangle
singularities~\cite{Nakamura:2019nch}. In 2003,
$Z_c(3900)$~\cite{Ablikim:2013mio,Liu:2013dau,Xiao:2013iha} and
$Z_c(4020)$~\cite{Ablikim:2013wzq} were observed. There are also
many model-depended interpretations of their inner structures, such
as the $D^{(*)}\bar{D}^{(*)}$ molecular
states~\cite{Liu:2008tn,Zhang:2013aoa,Aceti:2014kja,Chakrabarti:2014dna,Navarra:2001ju,Aceti:2014uea},
the $S$-wave tetraquark
states~\cite{Chen:2010ze,Zhao:2014qva,Faccini:2013lda,Deng:2015lca,Voloshin:2013dpa,Maiani:2007wz,Patel:2014vua},
and the kinematical effects
~\cite{Szczepaniak:2015eza,Chen:2011xk,Chen:2013coa,Swanson:2014tra,Swanson:2015bsa,Albaladejo:2015lob}.

Very recently, the BESIII Collaboration reported the first evidence
of the decay $Z_c(3900)^\pm\rightarrow \rho^\pm\eta_c$ with a
statistical significance of 3.9$\sigma$ in the
$\pi^+\pi^-\pi^0\eta_c$ final state~\cite{Ablikim:2019ipd}. The
BESIII Collaboration also gave the ratio between the partial widths
of the $\rho^{\pm}\eta_c$ and $\pi^{\pm}J/\psi$ decay modes at
$\sqrt{s}=4.226$ GeV~\cite{Ablikim:2019ipd}
\begin{eqnarray}
R^{\text{exp}}_{Z_c(3900)}=\frac{\mathcal{B}(Z_c(3900)^{\pm}\rightarrow
\rho^{\pm}\eta_c)}{\mathcal{B}(Z_c(3900)^{\pm}\rightarrow
\pi^{\pm}J/\psi)}=2.2\pm0.9.
\end{eqnarray}

Before the BESIII's measurement~\cite{Ablikim:2019ipd}, the relative
decay rate was predicted either in the molecular or tetraquark
scenarios within the framework of a covariant quark
model~\cite{Goerke:2016hxf}, the phenomenological Lagrangian field
theory~\cite{Patel:2014zja}, the nonrelativistic effective field
theory~\cite{Esposito:2014hsa}, the light front
model(LFM)~\cite{Ke:2013gia}, QCD sum
rules~\cite{Wang:2017lot,Dias:2013xfa,Agaev:2016dev}, etc. Later,
the author of Ref~\cite{Chen:2019wjd} studied the decay properties
of the $Z_c(3900)$ as a compact tetraquark state and a hadronic
molecular state through the Fierz rearrangement of the Dirac and
color indices. We collect the theoretical predictions in
Table~\ref{ratios1}, which differ greatly.
\begin{table}[h]
\caption{\label{ratios1} The theoretical predictions of
$R_{Z_c(3900)}$ in various models.}
\begin{tabular}{ccccccccccccccc}\hline\hline
Experiment~~~~&Molecular ~~~~&Tetraquark\\
\hline
$2.2\pm0.9$~\cite{Ablikim:2019ipd}~~~~&$0.046^{+0.025}_{-0.017}$~\cite{Esposito:2014hsa}~~~~&$230^{+330}_{-140}$~\cite{Esposito:2014hsa}\\
           ~~~~&$1.78^{+0.41}_{-0.37}$~\cite{Goerke:2016hxf}~~~~&$0.27^{+0.40}_{-0.17}$~\cite{Esposito:2014hsa}\\
           ~~~~&0.12~\cite{Ke:2013gia}~~~~&$1.86^{+0.41}_{-0.35}$~\cite{Goerke:2016hxf}\\
           ~~~~&0.007~\cite{Patel:2014zja}~~~~&$1.28^{+0.37}_{-0.30}$~\cite{Goerke:2016hxf}\\
           ~~~~&0.059~\cite{Chen:2019wjd} ~~~~&2.2~\cite{Chen:2019wjd}\\
           ~~~~&                          ~~~~&$1.08\pm0.88$~\cite{Wang:2017lot}\\
           ~~~~&                          ~~~~&$0.95\pm0.40$~\cite{Dias:2013xfa}\\
           ~~~~&                          ~~~~&0.66~\cite{Faccini:2013lda}\\
           ~~~~&                          ~~~~&$0.57\pm0.17$~\cite{Agaev:2016dev}\\
\hline\hline
\end{tabular}
\end{table}

In the present work we calculate the ratio between the $\rho\eta_c$
and $\pi J/\psi$ decay modes for the charged states $Z_c(3900)$,
$Z_c(4020)$ and $Z_c(4430)$ in two different scenarios. In scenario
I, we take the $Z_c(3900)$, $Z_c(4020)$ and $Z_c(4430)$ as the
$D\bar{D}^*$, $D^*\bar{D}^*$ and $D(2S)\bar{D}^*$ molecular states
with spin-parity $J^P=1^+$, respectively. In scenario II, we treat
the $Z_c(3900)$, $Z_c(4020)$ and $Z_c(4430)$ as the tetraquark
states. Our results show that in the molecular and tetraquark
scenarios, the ratios for $Z_c(3900)$ are
$R^{\text{th}}_{Z_c(3900)}\simeq1.3$ and
$R^{\text{th}}_{Z_c(3900)}\simeq1.6$, respectively, which are both
in the range of experimental result, $R^{\mathrm{exp}}=2.2\pm0.9$.
For $Z_c(4020)$, the ratio is about $R^{\text{th}}_{Z_c(4020)}\sim2$
in both two scenarios. As to the $Z_c(4430)$, we find that the ratio
in the molecular scenario ($R^{\text{th}}_{Z_c(4430)}\simeq1.4$) is
slightly smaller than that in the tetraquark scenario
($R^{\text{th}}_{Z_c(4430)}\simeq(1.7-1.4)$).

We notice that the ratios $\frac{\Gamma[Z_c(3900)\rightarrow
\rho\eta_c]}{\Gamma[Z_c(4020)\rightarrow \rho\eta_c]}$ and
$\frac{\Gamma[Z_c(3900)\rightarrow \pi
J/\psi]}{\Gamma[Z_c(4020)\rightarrow \pi J/\psi]}$ are about 12.5
and 24.2, respectively in the molecular scenario. However both
ratios are about 1.2 in the tetraquark scenario. In other words,
these ratios are very sensitive to the underlying dynamics of
$Z_c(3900)$ and $Z_c(4020)$, which may be helpful to pin down their
inner structures.

This paper is organized as follows. In Sec. II, we give an
introduction of the quark-exchange model and calculate the
transition amplitudes in the molecular and tetraquark scenarios.
Then we discuss and compare our results in the two scenarios in Sec.
III. We give a short summary in Sec. IV.

\section{Model introduction}
\subsection{Decay width}
For a four-quark state ($F$, for short) decaying into two particles
labelled as $C$ and $D$, the decay width in the rest frame of the
initial particle has the form
\begin{eqnarray}
d\Gamma=\frac{|\vec{p}_c|}{32\pi^2M^2}|\mathcal{M}(F\rightarrow
C+D)|^2d\Omega.
\end{eqnarray}
Here, $M$ represents the mass of the initial four-quark state $F$;
$\vec{p}_c$ denotes the three-momentum of the meson $C$ in the final
state; $\mathcal{M}(F\rightarrow C+D)$ is the transition amplitude
of the two-body decay $F\rightarrow C+D$, which is related to the
$T$-matrix via
\begin{eqnarray}
\mathcal{M}(F\rightarrow
C+D)=-(2\pi)^{3/2}\sqrt{2M}\sqrt{2E_C}\sqrt{2E_D}T,
\end{eqnarray}
where $E_C$ and $E_D$ denote the energy of the final mesons $C$ and
$D$, respectively. The $T$-matrix reads
\begin{eqnarray}
T&=&\langle\psi_{CD}(\vec{p}_c)|V_{\text{eff}}(\vec{k},\vec{p})|\psi_{F}(\vec{k})\rangle\nonumber\\
&=&\langle\psi_{CD}(\vec{p}_c)|V_{\text{eff}}(\vec{k},\vec{p})|\psi_{AB}(\vec{k})\rangle.
\end{eqnarray}
Here, $\psi_{CD}(\vec{p}_c)$ represents the relative spacial wave
function between the final mesons $C$ and $D$; $\psi_{AB}(\vec{k})$
is the normalized relative spacial wave function between the
constituent clusters $A$ and $B$. In molecular scenario, the
constituents represent mesons, while in tetraquark scenario the
constituents represent the diquark $[cq]$ and antidiquark
$[\bar{c}\bar{q}]$. $V_{\text{eff}}(\vec{k},\vec{p})$ denotes the
effective potential, which is in the general case a function of the
initial and final relative momentum $\vec{k}$ and $\vec{p}_c$.

The four-quark state may be a superposition of terms with different
orbital angular momenta~\footnote{We assume the orbital excitation
is between the diquark and antidiquark.}. Thus, the relative spacial
wave function in the momentum space has the form
\begin{eqnarray}
\psi_{AB}(\vec{k})=\sum_lR_l(k)Y_{lm}(\vec{k}).\label{relative}
\end{eqnarray}
Then, the Eq.~(4) can be written as
\begin{eqnarray}
T&=&\frac{1}{(2\pi)^3}\int d\vec{k}\int d\vec{p}\delta(\vec{p}-\vec{p}_c)V_{\text{eff}}(\vec{k},\vec{p})\sum_lR_l(k)Y_{lm}(\vec{k})\nonumber\\
&=&\frac{1}{(2\pi)^2}\sum_lM_{ll}Y_{lm}(\vec{p}_c),
\end{eqnarray}
where
\begin{eqnarray}
M_{ll}=\int_{-1}^{1}P_l(\mu)d\mu \int
dkV_{\text{eff}}(\vec{k},\vec{p}_c,\mu)R_l(k)k^2.
\end{eqnarray}
In this equation, $P_l(\mu)$ is Legendre function and $\mu$
represents the cosine of the angle between the
momenta $\vec{k}$ and $\vec{p}_c$.

Finally, with the relativistic phase space, the decay width of
two-body decay progress reads
\begin{eqnarray}
\Gamma_{\pi
J/\psi}=\frac{E_CE_D|\vec{p}_c|}{(2\pi)^3M}|M_{ll}|^2.\label{gama1}
\end{eqnarray}

For the $\rho\eta_c$ decay mode, we further consider the decay width
of the $\rho$ meson, and get
\begin{eqnarray}
\Gamma_{\rho\eta_c}=\frac{1}{N}\int
ds\frac{E_CE_D|\vec{p}_c|}{(2\pi)^3M}|M_{ll}|^2\frac{1}{\pi}\frac{m_{\rho}\Gamma_{\rho}}{(s-m^2_{\rho})^2+(m_{\rho},\Gamma_{\rho})^2}\label{gama2}
\end{eqnarray}
with
\begin{eqnarray}
N=\int
ds\frac{1}{\pi}\frac{m_{\rho}\Gamma_{\rho}}{(s-m^2_{\rho})^2+(m_{\rho}\Gamma_{\rho})^2}.
\end{eqnarray}
Here, $m_{\rho}$ and $\Gamma_{\rho}$ stand for the mass and total
decay width of the $\rho$ meson, respectively. $s$ denotes the
square of the $\rho$ meson invariant mass spectrum.

\subsection{Effective potential}
\subsubsection{The molecular scenario}
\begin{figure}[h]
\centering \epsfxsize=8 cm \epsfbox{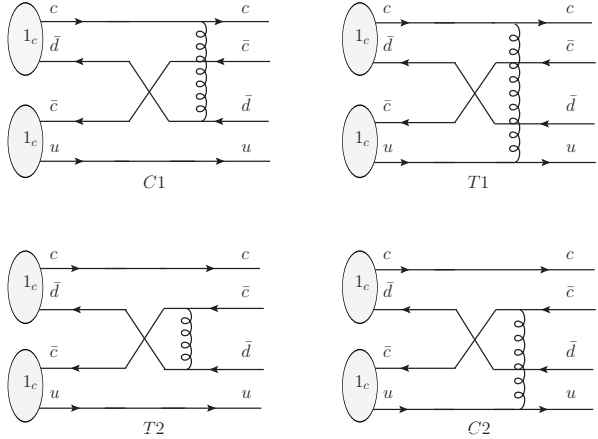} \caption{Diagrams for
the scattering process $AB\rightarrow CD$ in the molecular scenario.
}\label{post1}
\end{figure}

The $J^P$ quantum number of the $Z_c(3900)$, $Z_c(4020)$ and
$Z_c(4430)$ are $1^+$. In  the molecular scenario, we treat them as
the loosely bound $S$-wave $D\bar{D}^*$, $D^*\bar{D}^*$ and
$D(2S)\bar{D}^*$
 molecular states according to their mass spectra, respectively. At Born order, the effective potential $V_{\text{eff}}(\vec{k},\vec{p}_c,\mu)$ is related to the reacting amplitude of the meson-meson scattering process,
\begin{eqnarray}
A(12)+B(34)\rightarrow C(13)+D(24),
\end{eqnarray}
where 1(3) and 2(4) denote the $c$($\bar{c}$) quark and
$\bar{q}$($q$) quark, respectively. In the quark interchange
model~\cite{Barnes:1991em,Swanson:1992ec,Hilbert:2007hc,Barnes:1999hs,Barnes:2000hu},
the scattering Hamiltonian of the processes
$D^{(*)}/D(2S)+\bar{D}^*\rightarrow \eta_c(J/\psi)+\rho(\pi)$ is
estimated by the sum of the interactions between the inner quarks as
illustrated in Fig.~\ref{post1}. Moreover, the short-range
interactions are dominant in the scattering processes of two
open-charmed mesons into a ground charmonium state plus a
light-flavor meson. Thus, the scattering potential can be
approximated by the one-gluon-exchange (OGE) potential $V_{ij}$ at
quark level~\footnote{The interactions in Eq.~(\ref{Q3}) are the
Fourier transformation of the potential in Ref.~\cite{Wong:2001td}.
In the following. we perform our calculations in the momentum space
for the purpose of simplification. The constant potential in the
spatial space does not contribute due to the cancelation of the form
factors and we just omitted the term in Eq.~(\ref{Q3}).},
\begin{eqnarray}
V_{ij}=\frac{\lambda_i}{2}\frac{\lambda_j}{2}\left\{\frac{4\pi\alpha_s}{q^2}+\frac{6\pi
b}{q^4}-\frac{8\pi\alpha_s}{3m_im_j}\mathbf{s}_i\cdot\mathbf{s}_je^{-\frac{q^2}{4\sigma^2}}\right\}\label{Q3},
\end{eqnarray}
where $\lambda_i(\lambda^T_i)$ represents the quark (antiquark)
generator; $q$ is the transferred momentum; $b$ denotes the string
tension; $\sigma$ is the range parameter in the hyperfine spin-spin
interaction; $m_i~(m_j)$ and $\mathbf{s}_i~(\mathbf{s}_j)$
correspond to the interacting constituent quark mass and spin
operator; $\alpha_s$ is the running coupling constant,
\begin{eqnarray}
\alpha_s(Q^2)=\frac{12\pi}{(33-2n_f)\text{ln}(A+Q^2/B^2)}\label{Q4}.
\end{eqnarray}
In this equation, $Q^2$ is the square of the invariant masses of the
interacting quarks. The parameters in Eqs.~(\ref{Q3})-(\ref{Q4}) are
fitted by the mass spectra of the observed
mesons~\cite{Wong:2001td}, and their values are listed in
Table~\ref{parameters}.

\begin{table}[h]
\caption{\label{parameters} The parameters~\cite{Wong:2001td} used
in the quark model.}
\begin{tabular}{lllcccccccccccl}\hline\hline
Parameter~~&$b$~~~~~~~~&0.18~~~~GeV$^2$\\
         ~~&$\sigma$~~~~~~~~&0.897~~GeV\\
         ~~&$A$~~~~~~~~&10\\
        ~~&$B$~~~~~~~~&0.31~~~~GeV\\
\hline
Constituent quark mass~~&$m_q$~~~~~~~~&0.334~~GeV\\
                      ~~&$m_c$~~~~~~~~&1.776~~GeV\\
\hline\hline
\end{tabular}
\end{table}

In the quark model, the color-spin-flavor-space wave function for a
meson is
\begin{eqnarray}
\Psi=\omega_c\phi_f\chi_s\psi(\vec{p}),
\end{eqnarray}
where $\omega_c$, $\phi_f$, $\chi_s$ and $\psi(\vec{p})$ represent
the wave functions in the color, flavor, spin and momentum space,
respectively. Here, the wave functions of the mesons are determined
by fitting the mass spectra  in the Godfrey-Isgur
model~\cite{Godfrey:1985xj}.

According to the decomposition of the meson wave functions, the
effective potential can be given as the product of the factors,
\begin{eqnarray}
V_{\text{eff}}(\vec{k},\vec{p}_c,\mu)=I_{\text{color}}I_{\text{flavor}}I_{\text{spin-space}}.\label{effective}
\end{eqnarray}
Here, $I$ with the subscripts color, flavor and spin-space represent
the overlaps of the initial and final wave functions in the
corresponding space. The color factor $I_{\text{color}}$ reads
\begin{eqnarray}
I_{\text{color}}=\langle\omega^C_c(13)\omega^D_c(24)|\frac{\lambda_i}{2}\cdot\frac{\lambda_j}{2}|\omega^A_c(12)\omega^B_c(34)\rangle.
\end{eqnarray}
Its value in different diagrams in Fig.~\ref{post1} is listed in
Table~\ref{color}. For the flavor factor $I_{\text{flavor}}$, its
value is simply unity for all diagrams considered in this paper.

\begin{table}[h]
\caption{\label{color} The color factor $I_{\text{color}}$ within
the molecular scenario.}
\begin{tabular}{ccccccccccccccc}\hline\hline
~~~~&$12(C1)$~~~~&$14(T1)$~~~~&$32(T2)$~~~~&$34(C2)$\\
\hline
$I_{\text{color}}$~~~~&$-\frac{4}{9}$~~~~&$\frac{4}{9}$~~~~&$\frac{4}{9}$~~~~&$-\frac{4}{9}$\\
\hline\hline
\end{tabular}
\end{table}

For the $S$-wave decay process, the spin and space factors can be
decoupled. The spin factor $I_{\text{spin}}$ reads
\begin{eqnarray}
I_{\text{spin}}=\langle[\chi^C_s(13)\chi^D_s(24)]_{S'}|\hat{\mathcal{O}}_s|[\chi^A_s(12)\chi^B_s(34)]_S\rangle,
\end{eqnarray}
where $S(S')$ stands for the total spin of the initial(final)
system. The spin operator $\hat{\mathcal{O}}_s$ equals to unitary
for the Coulomb and linear interactions, and equals to
$\mathbf{s_i}\cdot \mathbf{s_j}$ for the spin-spin interaction. We
collect the values of color-spin factors $I_{\text{color}}\cdot
I_{\text{spin}}$ in Table~\ref{color-spin}.

\begin{table}[h]
\caption{\label{color-spin} The values of color-spin factors for the
diagrams $[C1,T1,T2,C2]$ within the molecular scenario. Here,
$D^{(*)}/D(2S)\bar{D}^{*}$ is the shorthand for
$D^{(*)}/D(2S)\bar{D}^{*}+c.c$.  }
\begin{tabular}{ccccccccccccccc}\hline\hline
\text{Initial state}~~&\text{Final state}~~&\text{Coul~\&~linear}~~&\text{Hyperfine}\\
\hline
$D\bar{D}^{*}$~~&$\eta_c\rho$~~&$\frac{2}{9}[-1,1,1,-1]$~~&$\frac{1}{18}[3,-1,3,-1]$\\
                 ~~&$J/\psi\pi$~~&$-\frac{2}{9}[-1,1,1,-1]$~~&$\frac{1}{18}[-3,-3,1,1]$\\
$D^{*}\bar{D}^{*}$~~&$\eta_c\rho$~~&$\frac{2\sqrt{2}}{9}[-1,1,1,-1]$
                 ~~&$-\frac{\sqrt{2}}{18}[1,1,1,1]$\\
                 ~~&$J/\psi\pi$~~&$\frac{2\sqrt{2}}{9}[-1,1,1,-1]$
                 ~~&$-\frac{\sqrt{2}}{18}[1,1,1,1]$\\
$D(2S)\bar{D}^{*}$~~&$\eta_c\rho$~~&$\frac{2}{9}[-1,1,1,-1]$~~&$\frac{1}{18}[3,-1,3,-1]$\\
                 ~~&$J/\psi\pi$~~&$-\frac{2}{9}[-1,1,1,-1]$~~&$\frac{1}{18}[-3,-3,1,1]$\\
\hline\hline
\end{tabular}
\end{table}

As to the spatial factor $I_{\text{space}}$, its expression reads
\begin{eqnarray}
I^{C1}_{\text{space}}&=&\int\int d\vec{q}d\vec{p}_3\psi_A(-\vec{q}-\vec{p}_3+\vec{p}_c-f_1\vec{k})\psi_B(\vec{p}_3+f_2\vec{k})\nonumber\\
&&\hat{\mathcal{O}}_q\psi^*_C(-\vec{p}_3+f_3\vec{p}_c)\psi^*_D(\vec{p}_3-f_4\vec{p}_c+\vec{k}),\label{space1}\\
I^{T1}_{\text{space}}&=&\int\int d\vec{q}d\vec{p}_3\psi_A(-\vec{q}-\vec{p}_3+\vec{p}_c-f_1\vec{k})\psi_B(\vec{p}_3+f_2\vec{k})\nonumber\\
&&\hat{\mathcal{O}}_q\psi^*_C(-\vec{p}_3+f_3\vec{p}_c)\psi^*_D(\vec{q}+\vec{p}_3-f_4\vec{p}_c+\vec{k}),\\
I^{T2}_{\text{space}}&=&\int\int d\vec{q}d\vec{p}_3\psi_A(-\vec{p}_3+\vec{p}_c-f_1\vec{k})\psi_B(\vec{p}_3+f_2\vec{k})\nonumber\\
&&\hat{\mathcal{O}}_q\psi^*_C(\vec{q}-\vec{p}_3+f_3\vec{p}_c)\psi^*_D(\vec{p}_3-f_4\vec{p}_c+\vec{k}),\\
I^{C2}_{\text{space}}&=&\int\int d\vec{q}d\vec{p}_3\psi_A(-\vec{q}-\vec{p}_3+\vec{p}_c-f_1\vec{k})\psi_B(\vec{p}_3+f_2\vec{k})\nonumber\\
&&\hat{\mathcal{O}}_q\psi^*_C(-\vec{q}-\vec{p}_3+f_3\vec{p}_c)\psi^*_D(\vec{q}+\vec{p}_3-f_4\vec{p}_c+\vec{k}).\label{space4}
\end{eqnarray}
In the equations, the spatial operator $\hat{\mathcal{O}}_q$
corresponds to $1/q^2$, $1/q^4$ and $e^{-q^2/(4\sigma^2)}$ for the
Coulomb, linear and spin-spin interactions, respectively. $\vec p_3$
denotes the momentum of the third quark. $f_i~(i=1,2,3,4)$ is a
constituent quark mass dependant function and expressed as
\begin{eqnarray}
&&f_1=\frac{m_1}{m_1+m_2}, ~~~~~~~~~~f_2=\frac{m_3}{m_3+m_4},\\
&&f_3=\frac{m_3}{m_1+m_3}, ~~~~~~~~~~f_4=\frac{m_4}{m_2+m_4}.
\end{eqnarray}
Here, $m_i~(i=1,2,3,4)$ represents the mass of the $i$-th quark.

Finally, with the obtained effective potential
$V_{\text{eff}}(\vec{k},\vec{p}_c,\mu)$, we can calculate the decay
widths by Eqs.~(\ref{gama1})-(\ref{gama2}) in cases that we know the
relative spacial wave function $\psi_{AB}(\vec{k})$ between mesons
$A$ and $B$. In the present work, we adopt an $S$-wave harmonic
ooscillator function to estimate the $S$-wave component of the
relative spacial wave function in Eq.~(\ref{relative}), which reads
\begin{eqnarray}
R_{00}(\vec{k})=\frac{2\text{exp}^{-\frac{k^2}{2\alpha^2}}}{\pi^{1/4}\alpha^{3/2}}.\label{relativeMOLECULAR}
\end{eqnarray}
The value of the harmonic oscillator strength $\alpha$ is related to
the root mean square radius $r_{\text{mean}}$ of the molecular state
by $\frac{\sqrt{3}}{\sqrt{2}\alpha}= r_{\text{mean}}$. Here, we take
the $r_{\text{mean}}$ in the range of (1.0-3.0)~fm, and the
corresponding value of $\alpha$ is collected in Table~\ref{alpha1}.

\begin{table}[h]
\caption{\label{alpha1} The corresponding values of the harmonic
oscillator strength $\alpha$ between the constituent mesons $A$ and
$B$ in the molecular scenario.}
\begin{tabular}{ccccccccccccccc}\hline\hline
$\sqrt{\langle r_{\text{mean}}\rangle^2}$~(fm)~~&1.0~~&1.2~~&1.5~~&1.7~~&2.0~~&2.4~~&3.0\\
\hline
$\alpha$~(GeV)~~&0.21~~&0.18~~&0.16~~&0.14~~&0.12~~&0.10~~&0.08\\
\hline\hline
\end{tabular}
\end{table}

\subsubsection{The tetraquark scenario}

For comparison, we further study the decays of the $Z_c(3900)$,
$Z_c(4020)$ and $Z_c(4430)$ as tetraquark states
$c\bar{c}q\bar{q}$~\cite{Maiani:2014aja},
\begin{eqnarray}
&&Z_c(3900):~~\frac{1}{\sqrt 2} \Big \{ \left[[cu]^{s=0}_{\bar{3}_c}[\bar{c}\bar{d}]^{s=1}_{3_c}\right]^{s=1}_{1_c}+\left[[cu]^{s=1}_{\bar{3}_c}[\bar{c}\bar{d}]^{s=0}_{3_c}\right]^{s=1}_{1_c}\Big \},\nonumber \\
&&Z_c(4020):~~\left[[cu]^{s=1}_{\bar{3}_c}[\bar{c}\bar{d}]^{s=1}_{3_c}\right]^{s=1}_{1_c},\nonumber \\
&&Z_c(4430):~~\frac{1}{\sqrt 2} \Big \{ \left[[cu]^{s=0}_{\bar{3}_c}[\bar{c}\bar{d}]^{s=1}_{3_c}\right]^{s=1}_{1_c}+\left[[cu]^{s=1}_{\bar{3}_c}[\bar{c}\bar{d}]^{s=0}_{3_c}\right]^{s=1}_{1_c}\Big \},\nonumber \\
\end{eqnarray}
where $Z_c(4430)$ is interpreted as the first radial excitation of
the $Z_c(3900)$.

\begin{figure}[h]
\centering \epsfxsize=8 cm \epsfbox{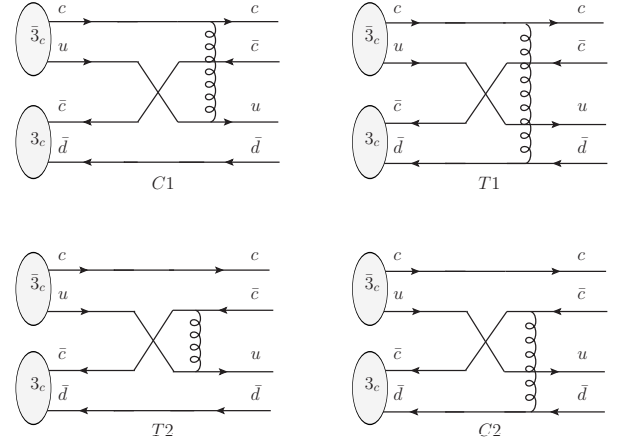} \caption{Diagrams for
the scattering process $AB\rightarrow CD$ in the tetraquark
scenario. }\label{post2}
\end{figure}

Similar to the molecular case, the
$V_{\text{eff}}(\vec{k},\vec{p}_c,\mu)$ can be approximated by the
interaction between the inner quarks, as shown in Fig.~\ref{post2}.

\begin{table*}[]
\caption{\label{color-spin-tetraquark} The values of the color-spin
factors for the diagrams $[C1,T1,T2,C2]$ within the tetraquark
scenario.}
\begin{tabular}{ccccccccccccccc}\hline\hline
\text{Initial state}~~~~~~~~&\text{Final state}~~~~~~~~&\text{Coul}~\&~\text{linear}~~~~~~~~&\text{Hyperfine}\\
\hline
$Z_c(3900)\big[[cu]^{S=0}_{\bar{3}_c}[\bar{c}\bar{d}]^{S=1}_{3_c}\big]^{S=1}_{1_c}$~~~~~~~~&$\eta_c\rho$~~~~~~~~&$\frac{1}{3\sqrt{3}}[-1,1,1,-1]$
                                                                 ~~~~~~~~&$[\frac{1}{4\sqrt{3}},-\frac{1}{12\sqrt{3}},\frac{1}{4\sqrt{3}},-\frac{1}{12\sqrt{3}}]$\\
                 ~~~~~~~~&$J/\psi\pi$~~~~~~~~&$-\frac{1}{3\sqrt{3}}[-1,1,1,-1]$
                 ~~~~~~~~&$[-\frac{1}{4\sqrt{3}},-\frac{1}{4\sqrt{3}},\frac{1}{12\sqrt{3}},\frac{1}{12\sqrt{3}}]$\\
$Z_c(4020)\big[[cu]^{S=1}_{\bar{3}_c}[\bar{c}\bar{d}]^{S=1}_{3_c}\big]^{S=1}_{1_c}$~~~~~~~~&$\eta_c\rho$~~~~~~~~&$\frac{2}{3\sqrt{6}}[-1,1,1,-1]$
                 ~~~~~~~~&$[-\frac{1}{6\sqrt{6}},-\frac{1}{6\sqrt{6}},-\frac{1}{6\sqrt{6}},-\frac{1}{6\sqrt{6}}]$\\
                 ~~~~~~~~&$J/\psi\pi$~~~~~~~~&$\frac{2}{3\sqrt{6}}[-1,1,1,-1]$
                 ~~~~~~~~&$[-\frac{1}{6\sqrt{6}},-\frac{1}{6\sqrt{6}},-\frac{1}{6\sqrt{6}},-\frac{1}{6\sqrt{6}}]$\\
$Z_c(4430)\big[[cu]^{S=0}_{\bar{3}_c}[\bar{c}\bar{d}]^{S=1}_{3_c}\big]^{S=1}_{1_c}$~~~~~~~~&$\eta_c\rho$~~~~~~~~&$\frac{1}{3\sqrt{3}}[-1,1,1,-1]$
                                                                 ~~~~~~~~&$[\frac{1}{4\sqrt{3}},-\frac{1}{12\sqrt{3}},\frac{1}{4\sqrt{3}},-\frac{1}{12\sqrt{3}}]$\\
                 ~~~~~~~~&$J/\psi\pi$~~~~~~~~&$-\frac{1}{3\sqrt{3}}[-1,1,1,-1]$
                 ~~~~~~~~&$[-\frac{1}{4\sqrt{3}},-\frac{1}{4\sqrt{3}},\frac{1}{12\sqrt{3}},\frac{1}{12\sqrt{3}}]$\\
\hline\hline
\end{tabular}
\end{table*}

The calculation of the $V_{\text{eff}}(\vec{k},\vec{p}_c,\mu)$ in
the tetraquark scenario is similar to that in the molecular
scenario. We can obtain the effective potential
$V_{\text{eff}}(\vec{k},\vec{p}_c,\mu)$ with Eq.~(\ref{effective})
as well. The flavor factor $I_{\text{flavor}}$ and spin factor
$I_{\text{spin}}$ are the same as those in the molecular scenario.
For the color factor $I_{\text{color}}$, there is a difference
between the two scenarios. In the molecular scenario, the initial
four-quark state is composed of two mesons, of which the color
configurations are $1_c$-$1_c$. However, in the tetraquark scenario,
the initial four-quark state is composed of diquark $[cq]$ and
antidiquark $[\bar{c}\bar{q}]$, of which the color configurations
are $3_c$-$\bar{3}_c$. The difference in color configurations may
result in quite different decay properties. The values of color-spin
factors are collected in Table~\ref{color-spin-tetraquark}.

To calculate the space factor $I_{\text{space}}$, we need the wave
function of the initial tetraquark state,
\begin{eqnarray}
\Psi(\vec{k}_r,\vec{k}_R,\vec{k}_X)=\psi_A(\vec{k}_r,\alpha_r)\psi_B(\vec{k}_R,\alpha_R)\psi_{AB}(\vec{k}_X,\alpha_X)~~~~~~~~~~~~\nonumber\\
\times[\chi_{s_a}(cu)\chi_{s_b}(\bar{c}\bar{d})]^{S_z}_S[\omega_{\bar{3}_c}(cu)\omega_{3_c}(\bar{c}\bar{d})]_{1_c}[\phi_{I_a}(cu)\phi_{I_b}(\bar{c}\bar{d})]^{I_z}_I,
\end{eqnarray}
where $\vec{k}_{r/R}$ denotes the momentum between the $c(\bar{c})$
and $u(\bar{d})$ quarks in the diquark (antidiquark), and
$\vec{k}_X$ is the one between the diquark $[cu]$ and antidiquark
$[\bar{c}\bar{d}]$. The $\alpha$ with the subscripts represents the
oscillating parameter along the corresponding Jacobi coordinates.

For the $Z_c({3900})$ and $Z_c(4020)$, the spatial wave function
$\psi$ is estimated by the $S$-wave harmonic oscillating wave
function,
\begin{eqnarray}
\psi(\vec{k},\alpha)=\frac{1}{\pi^{3/4}\alpha^{3/2}}\text{exp}^{-\frac{k^2}{2\alpha^2}}.
\end{eqnarray}
The $\alpha$ values are taken from Ref.~\cite{Deng:2015lca}, in
which the authors presented a systematic study of the tetraquark
states $[cu][\bar{c}\bar{d}]$ with the color flux-tube model, and
predicted that the charged charmonium-like states $Z_c(3900)$ could
be identified as the tetraquark state $[cu][\bar{c}\bar{d}]$ with
the quantum numbers $1^3S_1$ and $J^P=1^+$ as listed in
Table~\ref{tetraquark}. For $Z_c(4020)$, we give its wave function
via imitating the wave function of $Z_c(3900)$, listed in
Table~\ref{tetraquark} as well. It should be remarked that the spin
of the diquark $[cu]$ and antidiquark $[\bar{c}\bar{d}]$ both equal
to unitary for $Z_c(4020)$.

\begin{table}[h]
\caption{\label{tetraquark} The rms of the tetraquark states
$[cu][\bar{c}\bar{d}]$. $\sqrt{\langle r\rangle^2}$($\sqrt{\langle
R\rangle^2}$) denotes the distance between $c(\bar{c})$ and
$u(\bar{u})$ quarks; $\sqrt{\langle X\rangle^2}$ is the distance
between the diquark $[cu]$ and antidiquark $[\bar{c}\bar{u}]$; unit
of rms is fm.}
\begin{tabular}{lcccccccccccccc}\hline\hline
\text{States}~~&\text{tetraquark}~~&$\sqrt{\langle r\rangle^2}$~~&$\sqrt{\langle R\rangle^2}$~~&$\sqrt{\langle X\rangle^2}$\\
\hline
$Z_c(3900)$~\cite{Deng:2015lca}~~&$\big[[cu]^{S=0}_{\bar{3}_c}[\bar{c}\bar{u}]^{S=1}_{3_c}\big]^{S=1}_{1_c}$~~&0.90~~&0.90~~&0.48\\
$Z_c(4020)$~~&$\big[[cu]^{S=1}_{\bar{3}_c}[\bar{c}\bar{u}]^{S=1}_{3_c}\big]^{S=1}_{1_c}$~~&0.90~~&0.90~~&0.48\\
\hline\hline
\end{tabular}
\end{table}

As to $Z_c(4430)$, the spatial wave function of the diquark $[cu]$
is replaced by that of $D$, and the spatial wave function of
anti-diquark $[\bar{c}\bar{d}]$ is replaced by that of $D^*$. The
relative spatial wave function between the diquark $[cu]$ and
antidiquark $[\bar{c}\bar{d}]$ is estimated by an $2S$-wave harmonic
oscillating space-wave function
\begin{eqnarray}
R_{10}(k_X)=\frac{\sqrt{6}\text{exp}^{-\frac{k^2_X}{2\alpha_X^2}}}{\pi^{1/4}\alpha_X^{3/2}}(1-\frac{2k^2_X}{3\alpha_X^2}).
\end{eqnarray}
The value of the harmonic oscillator strength $\alpha_X$ is related
to the root mean square radius $r_{X}$ of the tetraquark state by
$\frac{\sqrt{7}}{\sqrt{2}\alpha_X}= r_{X}$. We vary the $r_{X}$ in
the rang of (0.5-2.0)~fm and the corresponding value of $\alpha_X$
is listed in Table~\ref{alphaX}.

\begin{table}[h]
\caption{\label{alphaX} The corresponding values of the harmonic
oscillator strength $\alpha_X$ between the diquark $[cu]$ and
antidiquark $[\bar{c}\bar{d}]$ for $Z_c(4430)$ as a tetraquark
state.}
\begin{tabular}{ccccccccccccccc}\hline\hline
$\sqrt{\langle r_X\rangle^2}$~(fm)~~~~~&0.5~~~~~&0.8~~~~~&1.0~~~~~&1.5~~~~~&2.0\\
\hline
$\alpha_X$~(GeV)~~~~~&0.74~~~~~&0.46~~~~~&0.37~~~~~&0.25~~~~~&0.18\\
\hline\hline
\end{tabular}
\end{table}

\section{Results}

Inspired by the recent measurement of the decay
$Z_c(3900)^{\pm}\rightarrow \rho^{\pm}\eta_c$ by the BESIII
Collaboration, we calculate the ratios between the $\rho\eta_c$ and
$\pi J/\psi$ decay modes for the charged states $Z_c(3900)$,
$Z_c(4020)$ and $Z_c(4430)$ in the molecular and tetraquark
scenarios. Our results and theoretical predictions are presented as
follows.

\subsection{The molecular scenario}
\begin{figure}[h]
\centering \epsfxsize=5.5 cm \epsfbox{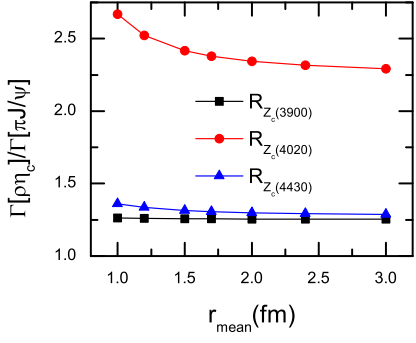} \caption{The
variation of the partial decay width ratios between the $\eta_c\rho$
and $J/\psi\pi$ channels for the $Z_c(3900)$, $Z_c(4020)$ and
$Z_c(4430)$ as the $D\bar{D}^{*}$, $D^*\bar{D}^{*}$ and
$D(2S)\bar{D}^{*}$ molecular states, respectively. Their masses are
fixed respectively on physical masses, namely 3886.6 MeV, 4024.1 MeV
and 4478 MeV.}\label{ratios}
\end{figure}

The mass of $Z_c(3900)$ ($M=3886.6\pm2.4$ MeV) is slightly higher
than the mass threshold of the $D\bar{D}^*$($\sim$3872 MeV). In the
molecular scenario, we take the $Z_c(3900)$ as a $D\bar{D}^{*}$
resonance molecular state, and calculate its branching fraction
ratio between the $\rho\eta_c$ and $\pi J/\psi$ decay modes.
Considering the uncertainty of the effective size for the molecular
state, we plot the ratio as a function of the root mean square
radius $r_{\text{mean}}$ in Fig.~\ref{ratios}. The ratio is
\begin{eqnarray}
R^{\text{th}}_{Z_c(3900)}\sim1.3,
\end{eqnarray}
which roughly accords with the experiment result
$R^{\text{exp}}_{Z_c(3900)}=2.2\pm0.9$~\cite{Ablikim:2019ipd} within
errors. Meanwhile the ratio is insensitive to $r_{\text{mean}}$ in
the range of (1.0$\sim$3.0) fm we considered in this work.

\begin{figure*}[]
\centering \epsfxsize=16 cm \epsfbox{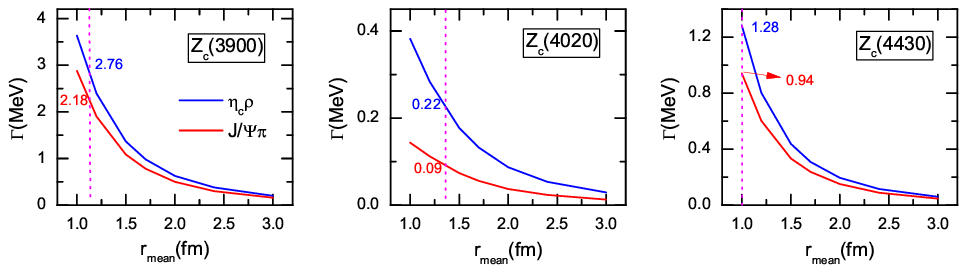} \caption{The
partial widths of the $\eta_c\rho$ and $J/\psi\pi$ decay modes for
$Z_c(3900)$, $Z_c(4020)$ and $Z_c(4430)$ as the $D\bar{D}^{*}$,
$D^*\bar{D}^{*}$ and $D(2S)\bar{D}^{*}$ molecular states,
respectively. Their masses are fixed respectively on the physical
masses, namely 3886.6 MeV, 4024.1 MeV and 4478 MeV.}\label{widths}
\end{figure*}

With the estimated relative spacial wave function as illustrated in
Eq.~(\ref{relativeMOLECULAR}), we further obtain the partial widths
of the $\eta_c\rho$ and $ J/\psi\pi$ decay modes and show them in
Fig.~\ref{widths}. It is obvious that the partial widths are
sensitive to $r_{\text{mean}}$ and vary from one MeV to
$\mathcal{O}(10^{-2})$ MeV. With $r_{\text{mean}}$ increasing, the
partial decay widths become smaller, or even close to zero. This can
be easily understood since the larger $r_{\text{mean}}$ means the
freer mesons $A$ and $B$. It is more difficult to interact with each
other and the effective potential tends to vanish.

Moreover, for an $S$-wave molecule composed of two mesons $A$ and
$B$, its size may be estimated
by~\cite{Guo:2017jvc,Weinberg:1962hj,Weinberg:1963zza}
\begin{eqnarray}
r_{\text{mean}}\sim1/\sqrt{2\mu|m_A+m_B-M|},\label{typicalsize}
 \end{eqnarray}
with the reduced mass $\mu=\frac{m_Am_B}{m_A+m_B}$. Then, the
typical size of $Z_c(3900)$ is estimated to be
$r_{\text{mean}}\simeq1.14$ fm. Hence we obtain
\begin{eqnarray}
&&\Gamma[Z_c(3900)\rightarrow\eta_c\rho]\sim2.76~\text{MeV},\nonumber\\
&&\Gamma[Z_c(3900)\rightarrow J/\psi\pi]\sim2.18~\text{MeV},
\end{eqnarray}
for $Z_c(3900)$ with a mass of $M=3886.6$ MeV (see
Table~\ref{decaywidths}).

\begin{table}[h]
\caption{\label{decaywidths} The partial decay widths (MeV) for the
$Z_c(3900)$, $Z_c(4020)$ and $Z_c(4430)$ as the molecular states
with typical size $r_{\text{mean}}$ (fm). $R^{\text{th}}$ and
$R^{\text{exp}}$ are theoretical and experimental ratios,
respectively.}
\begin{tabular}{ccccccccccccccc}\hline\hline
state~~~~~&$r_{\text{mean}}$~~~~~&$\Gamma[\eta_c\rho]$~~~~~&$\Gamma[J/\psi\pi]$~~~~~&$R^{\text{th}}$~~~~~&$R^{\text{exp}}$\\
\hline
$Z_c(3900)$~~~~~&1.14~~~~~&2.76~~~~~&2.18~~~~~&1.3~~~~~&$2.2\pm0.9$\\
$Z_c(4020)$~~~~~&1.37~~~~~&0.22~~~~~&0.09~~~~~&2.4~~~~~&$\cdot\cdot\cdot$\\
$Z_c(4430)$~~~~~&1.00~~~~~&1.28~~~~~&0.94~~~~~&1.4~~~~~&$\cdot\cdot\cdot$\\
\hline\hline
\end{tabular}
\end{table}

For $Z_c(4020)$, we take it as the $S$-wave $D^*\bar{D}^*$ resonance
molecular state since its mass ($M$=4024.1 MeV) is slightly about 10
MeV higher than the mass threshold of the $D^*\bar{D}^*$. With the
molecular size varying in the range of
$r_{\text{mean}}$=(1.0$\sim$3.0) fm, we calculate its partial decay
width ratio between the $\eta_c\rho$ and $ J/\psi\pi$ modes, and
obtain
 \begin{eqnarray}
R^{\text{th}}_{Z_c(4020)}\sim(2.7\sim2.3),
\end{eqnarray}
with the mass being $M$=4024.1 MeV (see Fig.~\ref{ratios}). This
value is almost independent of $r_{\text{mean}}$ we considered in
the present work.

We also plot the partial decay widths of the $\eta_c\rho$ and $
J/\psi\pi$ modes versus the molecular size $r_{\text{mean}}$ in
Fig.~\ref{widths}. In the figure, we find that the partial widths
are about $\mathcal{O}(10^{-1}\sim10^{-2})$ MeV, and strongly
dependent on $r_{\text{mean}}$. Fixing $r_{\text{mean}}\simeq1.37$
fm estimated by Eq.~(\ref{typicalsize}), we obtain
\begin{eqnarray}
&&\Gamma[Z_c(4020)\rightarrow\eta_c\rho]\sim0.22~\text{MeV},\nonumber\\
&&\Gamma[Z_c(4020)\rightarrow J/\psi\pi]\sim0.09~\text{MeV}.
\end{eqnarray}
The predicted branching ratios are
\begin{eqnarray}
&&\mathcal{B}[Z_c(4020)\rightarrow \eta_c\rho]\sim1.7\%,\nonumber\\
&&\mathcal{B}[Z_c(4020)\rightarrow J/\psi\pi]\sim0.7\%.
\end{eqnarray}
which are quite small.

The partial widths of the $\eta_c\rho$ and $ J/\psi\pi$ decay modes
for $Z_c(4020)$ are smaller than those for $Z_c(3900)$,
\begin{eqnarray}
&&\frac{\Gamma[Z_c(3900)\rightarrow\eta_c\rho]}{\Gamma[Z_c(4020)\rightarrow\eta_c\rho]}=12.5,\label{molecularratios1}\\
&&\frac{\Gamma[Z_c(3900)\rightarrow
J/\psi\pi]}{\Gamma[Z_c(4020)\rightarrow J/\psi
\pi]}=24.2.\label{molecularratios2}
\end{eqnarray}
This  indicates that the couplings of the $D\bar{D}^*$ to the
$\eta_c\rho$ and $ J/\psi\pi$ channels are stronger than those of
the $D^*\bar{D}^*$. The main difference between the $D\bar{D}^*$ and
$D^*\bar{D}^*$ is the spin wave function. Our results show that in
the molecular scenario, different spin-spin coupling may have a
great impact on the strong decay properties. We take the $J/\psi
\pi$ decay mode as an example. In Table \ref{color-spin}, the spin
factor for the coupling with the  $D\bar{D}^*$  is three times
larger than that of the $D^*\bar{D}^*$ in Fig.~\ref{post1}-$C1$ and
Fig.~\ref{post1}-$T1$. The hyperfine interaction is expected to be
more important for the $J/\psi \pi$ decay mode of $Z_c(3900)$.
Moreover, our calculation shows that the hyperfine interaction for
the $Z_c(3900)$ plays a quite important role in
Fig.~\ref{post1}-$C1$ and even change the sign of its amplitude.

As to $Z_c(4430)$, in molecular scenario, we take it as an $S$-wave
$D(2S)\bar{D}^*$ molecular state. Similarly we change the size of
the molecular state in the range of $r_{\text{mean}}$=(1.0$\sim$3.0)
fm, and obtain
\begin{eqnarray}
R^{\text{th}}_{Z_c(4430)}\sim(1.4\sim1.3),
\end{eqnarray}
for $Z_c(4430)$ with a mass of $M$=4478 MeV (see Fig.~\ref{ratios}).
Meanwhile, the partial widths of the $\eta_c\rho$ and $ J/\psi\pi$
modes as the function of $r_{\text{mean}}$ for $Z_c(4430)$ are shown
in Fig.~\ref{widths} as well. According to the figure, the decay
properties of the $D(2S)\bar{D}^*$ molecular state are similar to
the $D^*\bar{D}^*$ molecular state.

Fixing $r_{\text{mean}}\simeq1.00$ fm, we further obtain
\begin{eqnarray}
&&\Gamma[Z_c(4430)\rightarrow\eta_c\rho]\sim1.28~\text{MeV},\nonumber\\
&&\Gamma[Z_c(4430)\rightarrow J/\psi\pi]\sim0.94~\text{MeV}.
\end{eqnarray}
At present, the charged state $Z_c(4430)$ was observed both in the
$\psi'\pi^{\pm}$ and $J/\psi\pi^{\pm}$
channels~\cite{Choi:2007wga,Aubert:2008aa,Mizuk:2009da,Chilikin:2013tch,Chilikin:2014bkk,Aaij:2014jqa},
and has not been reported in the $\eta_c\rho$ channel. According to
our theoretical predictions, if $Z_c(4430)$ is a $D(2S)\bar{D}^*$
molecular state, the partial width of $\eta_c\rho$ is comparable to
that of $J/\psi\pi$, which indicates this state may be observed in
the $\eta_c\rho$ channel as well.

So far, we have obtained the decay ratios in the molecular scenario.
We find that the ratios are not sensitive to the relative molecular
wave function in the loosely bound system while the partial decay
widths are very sensitive to the size of the molecules because of
the sensitivity of the effective potentials.

\subsection{The tetraquark scenario}

In the tetraquark scenario, we obtain the decay ratio for the
$Z_c(3900)$ state
\begin{eqnarray}
R^{\text{th}}_{Z_c(3900)}\sim1.6,
\end{eqnarray}
which agrees with the experimental result (see
table~\ref{decaywidthstetra}).

\begin{table}[h]
\caption{\label{decaywidthstetra} The partial decay widths (MeV) for
the $Z_c(3900)$ and $Z_c(4020)$ as the tetraquark states.
$R^{\text{th}}$ and $R^{\text{exp}}$ are the theoretical and
experimental ratios, respectively.}
\begin{tabular}{ccccccccccccccc}\hline\hline
state~~~~~&$\Gamma[\eta_c\rho]$~~~~~&$\Gamma[J/\psi\pi]$~~~~~&$R^{\text{th}}$~~~~~&$R^{\text{exp}}$\\
\hline
$Z_c(3900)$~~~~~&0.23~~~~~&0.14~~~~~&1.6~~~~~&$2.2\pm0.9$\\
$Z_c(4020)$~~~~~&0.19~~~~~&0.12~~~~~&1.6~~~~~&$\cdot\cdot\cdot$\\
\hline\hline
\end{tabular}
\end{table}

The predicted partial decay widths of the $\eta_c\rho$ and $
J/\psi\pi$ modes are
\begin{eqnarray}
&&\Gamma[Z_c(3900)\rightarrow\eta_c\rho]\sim0.23~\text{MeV},\nonumber\\
&&\Gamma[Z_c(3900)\rightarrow J/\psi\pi]\sim0.14~\text{MeV}.
\end{eqnarray}

Via imitating the wave function of $Z_c(3900)$, we estimate the wave
function of $Z_c(4020)$ as listed in Table~\ref{tetraquark}.
Similarly we fix the mass of $Z_c(4020)$ at $M=4024.1$ MeV and
obtain
\begin{eqnarray}
&&\Gamma[Z_c(4020)\rightarrow\eta_c\rho]\sim0.19~\text{MeV},\nonumber\\
&&\Gamma[Z_c(4020)\rightarrow J/\psi\pi]\sim0.12~\text{MeV}.
\end{eqnarray}
Then the predicted partial decay widths ratio is
\begin{eqnarray}
R^{\text{th}}_{Z_c(4020)}\sim1.6.
\end{eqnarray}
The decay properties of $Z_c(4020)$ are similar in the molecular and
tetraquark scenarios. Thus, besides the decay ratios, more precise
experimental information is required to pin down the inner structure
of this state.

As shown in table~\ref{decaywidthstetra}, the partial widths of the
$\eta_c\rho$ and $J/\psi\pi$ decay modes for $Z_c(4020)$ are
comparable to those for $Z_c(3900)$,
\begin{eqnarray}
&&\frac{\Gamma[Z_c(3900)\rightarrow\eta_c\rho]}{\Gamma[Z_c(4020)\rightarrow\eta_c\rho]}=1.2,\\
&&\frac{\Gamma[Z_c(3900)\rightarrow
J/\psi\pi]}{\Gamma[Z_c(4020)\rightarrow J/\psi \pi]}=1.2.
\end{eqnarray}
The ratios are very different from those in
Eqs.~(\ref{molecularratios1})-(\ref{molecularratios2}). As mentioned
earlier, in the molecular scenario the hyperfine interaction plays a
quite important role for $Z_c(3900)$ in Fig.2-$C1$ and even changes
the sign of its amplitude. Thus the total amplitudes for $Z_c(3900)$
are much larger than those for $Z_c(4020)$. However, in the
tetraquark scenario the Coulomb and linear interactions are dominant
for both states $Z_c(3900)$ and $Z_c(4020)$. There exists good
evidence for $Z_c(3900)$ in the $\eta_c\rho$ and $J/\psi\pi$
channels experimentally and no evidence for $Z_c(4020)$. Our results
support that the two states are more likely to be the molecular
states.

\begin{figure}[h]
\centering \epsfxsize=8 cm \epsfbox{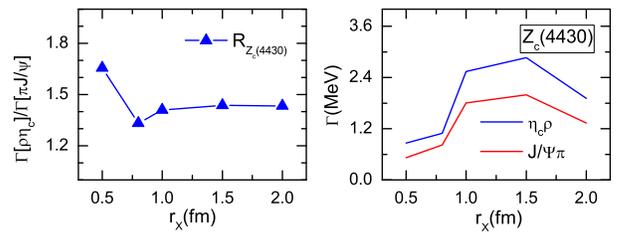}
\caption{Left (a): the branching fraction ratio between $\eta_c\rho$
and $J/\psi\pi$ for $Z_c(4430)$ in tetraquark scenario. Right (b):
the partial decay widths for the $Z_c(4430)$  decaying into the
$\eta_c\rho$ and $J/\psi\pi$ channels  in tetraquark scenario.
}\label{4430tetraquark}
\end{figure}

For the $Z_c(4430)$, with the estimated wave function we plot the
partial decay width ratio between the $\eta_c\rho$ and $J/\psi\pi$
decay modes as a function of the effective size $r_X$ of the
tetraquark state (see Fig.~\ref{4430tetraquark}). We find that the
ratio slightly depends on $r_X$. Varying the $r_X$ in the range
$r_X=(0.5-2.0)$ fm, the ratio is
\begin{eqnarray}
R^{\text{th}}_{Z_c(4430)}\sim(1.7\sim1.4),
\end{eqnarray}
which is slightly larger than that as a molecular state. According
to our results, the branching fraction ratio between the
$\eta_c\rho$ and $J/\psi\pi$ modes of the $Z_c(4430)$ as a molecule
or a tetraquark state is larger than one, which indicates that the
$Z_c(4430)$ is more easier to decay into the $\eta_c\rho$ channel.

So far, we have calculated the decay ratios of the $Z_c(3900)$,
$Z_c(4020)$ and $Z_c(4430)$ decaying into the $\eta_c\rho$ and
$J/\psi\pi$ channels in the molecular and tetraquark scenarios. Our
results show that the decay ratios in both scenarios are similar to
each other. We cannot determine the inner structures only with the
decay ratios. However, if we look at the partial decay widths, we
find that in molecular scenario, the $\Gamma(Z_c(3900)\rightarrow
J/\psi\pi)$ are much larger than the $\Gamma(Z_c(4020)\rightarrow
J/\psi\pi)$, while they are similar in the tetraquark scenario. In
experiments, the $Z_c(3900)$ state is observed in the $J/\psi\pi$
invariant mass spectrum while no significant $Z_c(4020)$ signal is
observed. This may support the states $Z_c(3900)$ and $Z_c(4020)$ as
the molecules instead of tightly bound tetraquark states.

\section{Summary}

In the present work, we calculate the branching fraction ratios
between the $\eta_c\rho$ and $J/\psi\pi$ decay modes for the charged
states $Z_c(3900)$, $Z_c(4020)$ and $Z_c(4430)$ with a quark
interchange model. In order to compare the decay properties in
different physical scenarios and pin down the inner structure of
these three mysterious charmonium-like states, we study the ratios
in the molecular and tetraquark scenarios, respectively. Meanwhile,
we estimate the absolute partial decay widths for the $\eta_c\rho$
and $J/\psi\pi$ decay channels. Our main results are summarized as
follows.

For $Z_c(4430)$, the branching fraction ratio as an S-wave
$D(2S)\bar D^*$ molecule ($R^{\text{th}}_{Z_c(4430)}\simeq1.4$) is
slightly smaller than that in the tetraquark scenario
($R^{\text{th}}_{Z_c(4430)}\simeq1.7\sim1.4$). We notice that the
ratios in both two physical scenarios are larger than 1, which
indicates that the $Z_c(4430)$ prefers to decay into the
$\rho\eta_c$ channel rather than the $\pi J/\psi$ channel. Besides
the $\pi J/\psi$ channel, the $\rho\eta_c$ may be another
interesting channel for the observation of $Z_c(4430)$ in future
experiments.

For $Z_c(3900)$, we obtain that the ratios are
$R^{\text{th}}_{Z_c(3900)}\sim1.3$ and $1.6$ in the molecular and
tetraquark scenarios, respectively. Both are comparable with the
experimental result. For $Z_c(4020)$, the ratios are
$R^{\text{th}}_{Z_c(4020)}\sim2.4$ and $1.6$, respectively. The
above results show that the ratios in both scenarios are similar to
each other. Thus, to investigate the inner structures, considering
only the decay ratio $R$ of the $Z_c$ itself is not enough.

In the molecular scenario, the partial decay widths of the
$\eta_c\rho$ and $ J/\psi\pi$ modes for $Z_c(4020)(D^*\bar{D}^*)$
are smaller than those for $Z_c(3900)(D\bar{D}^*)$ by one order. In
the molecular scenario, different spin-spin coupling may have a
great impact on the strong decay properties. On the other side, the
partial decay widths of the $\eta_c\rho$ and $ J/\psi\pi$ modes for
$Z_c(4020)(D^*\bar{D}^*)$ are comparable to those for $Z_c(3900)$ in
the tetraquark scenario. At present, there exists good evidence for
$Z_c(3900)$ in the $\eta_c\rho$ and $J/\psi\pi$ channels
experimentally and no evidence for $Z_c(4020)$. Our results indicate
that these two states are more likely to be the molecule-like states
which arise from the $D^{(*)}{\bar D}^{(*)}$ hadronic interactions.

\section*{Acknowledgements }

We would like to thank Xiao-Lin Chen and Wei-Zhen Deng for very
helpful suggestions. This work is supported by the National Natural
Science Foundation of China under Grants No.11947048 and 11975033.
G. J. Wang is supported by China Postdoctoral Science Foundation
No.2019M660279.



\begin{thebibliography}{99}
\bibitem{Choi:2003ue}
  S.~K.~Choi {\it et al.} [Belle Collaboration],
  Observation of a narrow charmonium - like state in exclusive $B^{+-} \rightarrow K^{+-} \pi^+ \pi^- J/\psi$ decays,
  Phys.\ Rev.\ Lett.\  {\bf 91}, 262001 (2003).

\bibitem{Tanabashi:2018oca}
  M.~Tanabashi {\it et al.} [Particle Data Group],
  Review of Particle Physics,
  Phys.\ Rev.\ D {\bf 98}, 030001 (2018).

\bibitem{Chen:2016qju}
  H.~X.~Chen, W.~Chen, X.~Liu and S.~L.~Zhu,
  The hidden-charm pentaquark and tetraquark states,
  Phys.\ Rept.\  {\bf 639}, 1 (2016).

\bibitem{Liu:2019zoy}
  Y.~R.~Liu, H.~X.~Chen, W.~Chen, X.~Liu and S.~L.~Zhu,
  Pentaquark and Tetraquark states,
  Prog.\ Part.\ Nucl.\ Phys.\  {\bf 107}, 237 (2019).

\bibitem{Brambilla:2019esw}
  N.~Brambilla, S.~Eidelman, C.~Hanhart, A.~Nefediev, C.~P.~Shen, C.~E.~Thomas, A.~Vairo and C.~Z.~Yuan,
  The $XYZ$ states: experimental and theoretical status and perspectives,
  arXiv:1907.07583 [hep-ex].

\bibitem{Guo:2017jvc}
  F.~K.~Guo, C.~Hanhart, U.~G.~Meiner, Q.~Wang, Q.~Zhao and B.~S.~Zou,
  Hadronic molecules,
  Rev.\ Mod.\ Phys.\  {\bf 90}, 015004 (2018).

\bibitem{Esposito:2016noz}
  A.~Esposito, A.~Pilloni and A.~D.~Polosa,
  Multiquark Resonances,
  Phys.\ Rept.\  {\bf 668}, 1 (2017).

\bibitem{Hosaka:2016pey}
  A.~Hosaka, T.~Iijima, K.~Miyabayashi, Y.~Sakai and S.~Yasui,
  Exotic hadrons with heavy flavors: X, Y, Z, and related states,
  PTEP {\bf 2016}, 062C01 (2016).
  [arXiv:1603.09229 [hep-ph]].

\bibitem{Choi:2007wga}
  S.~K.~Choi {\it et al.} [Belle Collaboration],
  Observation of a resonance-like structure in the $\pi^{\pm} \psi^{\prime}$ mass distribution in exclusive $B \to K \pi^{\pm} \psi^{\prime}$ decays,
  Phys.\ Rev.\ Lett.\  {\bf 100}, 142001 (2008).

\bibitem{Aubert:2008aa}
  B.~Aubert {\it et al.} [BaBar Collaboration],
  Search for the $Z(4430)^-$ at BABAR,
  Phys.\ Rev.\ D {\bf 79}, 112001 (2009).

\bibitem{Mizuk:2009da}
  R.~Mizuk {\it et al.} [Belle Collaboration],
  Dalitz analysis of $B\rightarrow K \pi^+ \psi^{\prime}$ decays and the $Z(4430)^+$,
  Phys.\ Rev.\ D {\bf 80}, 031104 (2009).

\bibitem{Chilikin:2013tch}
  K.~Chilikin {\it et al.} [Belle Collaboration],
  Experimental constraints on the spin and parity of the $Z$(4430)$^+$,
  Phys.\ Rev.\ D {\bf 88}, 074026 (2013).

\bibitem{Chilikin:2014bkk}
  K.~Chilikin {\it et al.} [Belle Collaboration],
  Observation of a new charged charmoniumlike state in $\bar{B}^0 \rightarrow J/\psi K^-\pi^+$ decays,
  Phys.\ Rev.\ D {\bf 90}, 112009 (2014).

\bibitem{Aaij:2014jqa}
  R.~Aaij {\it et al.} [LHCb Collaboration],
  Observation of the resonant character of the $Z(4430)^-$ state,
  Phys.\ Rev.\ Lett.\  {\bf 112}, 222002 (2014).

\bibitem{Ma:2014zua}
  L.~Ma, X.~H.~Liu, X.~Liu and S.~L.~Zhu,
  Exotic Four Quark Matter: $Z_1(4475)$,
  Phys.\ Rev.\ D {\bf 90}, 037502 (2014).


\bibitem{Maiani:2007wz}
  L.~Maiani, A.~D.~Polosa and V.~Riquer,
  The Charged Z(4433): Towards a new spectroscopy,
  arXiv:0708.3997 [hep-ph].

\bibitem{Ebert:2008kb}
  D.~Ebert, R.~N.~Faustov and V.~O.~Galkin,
  Excited heavy tetraquarks with hidden charm,
  Eur.\ Phys.\ J.\ C {\bf 58}, 399 (2008).

\bibitem{Maiani:2014aja}
  L.~Maiani, F.~Piccinini, A.~D.~Polosa and V.~Riquer,
  The $Z(4430)$ and a New Paradigm for Spin Interactions in Tetraquarks,
  Phys.\ Rev.\ D {\bf 89}, 114010 (2014).


\bibitem{Bugg:2008wu}
  D.~V.~Bugg,
  How Resonances can synchronise with Thresholds,
  J.\ Phys.\ G {\bf 35}, 075005 (2008).

\bibitem{Nakamura:2019nch}
  S.~X.~Nakamura,
  $Z_c(4430)$, $Z_c(4200)$, $Z_1(4050)$, and $Z_2(4250)$ as triangle singularities,
  arXiv:1909.03976 [hep-ph].





\bibitem{Ablikim:2013mio}
  M.~Ablikim {\it et al.} [BESIII Collaboration],
  Observation of a Charged Charmoniumlike Structure in $e^+e^-\rightarrow \pi^+\pi^- J/\psi$ at $\sqrt{s}$ =4.26  GeV,
  Phys.\ Rev.\ Lett.\  {\bf 110}, 252001 (2013).

\bibitem{Liu:2013dau}
  Z.~Q.~Liu {\it et al.} [Belle Collaboration],
  Study of $e^+e^-\rightarrow \pi^+\pi^- J/\psi$ and Observation of a Charged Charmoniumlike State at Belle,
  Phys.\ Rev.\ Lett.\  {\bf 110}, 252002 (2013).

\bibitem{Xiao:2013iha}
  T.~Xiao, S.~Dobbs, A.~Tomaradze and K.~K.~Seth,
  Observation of the Charged Hadron $Z_c^{\pm}(3900)$ and Evidence for the Neutral $Z_c^0(3900)$ in $e^+e^-\to \pi\pi J/\psi$ at $\sqrt{s}=4170$ MeV,
  Phys.\ Lett.\ B {\bf 727}, 366 (2013).


\bibitem{Ablikim:2013wzq}
  M.~Ablikim {\it et al.} [BESIII Collaboration],
  Observation of a Charged Charmoniumlike Structure $Z_c$(4020) and Search for the $Z_c$(3900) in $e^+e^- \rightarrow \pi^+\pi^-h_c$,
  Phys.\ Rev.\ Lett.\  {\bf 111}, 242001 (2013).

\bibitem{Liu:2008tn}
  X.~Liu, Z.~G.~Luo, Y.~R.~Liu and S.~L.~Zhu,
  $X(3872)$ and Other Possible Heavy Molecular States,
  Eur.\ Phys.\ J.\ C {\bf 61}, 411 (2009).

\bibitem{Aceti:2014kja}
  F.~Aceti, M.~Bayar, J.~M.~Dias and E.~Oset,
  Prediction of a $Z_c(4000)$ $D^* \bar D^*$ state and relationship to the claimed $Z_c(4025)$,
  Eur.\ Phys.\ J.\ A {\bf 50}, 103 (2014).

\bibitem{Zhang:2013aoa}
  J.~R.~Zhang,
  Improved QCD sum rule study of $Z_{c}(3900)$ as a $\bar{D}D^{*}$ molecular state,
  Phys.\ Rev.\ D {\bf 87}, 116004 (2013).


\bibitem{Chakrabarti:2014dna}
  D.~Chakrabarti and C.~Mondal,
  Transverse charge and magnetization densities in holographic QCD,
  Eur.\ Phys.\ J.\ C {\bf 74}, 2962 (2014).

\bibitem{Navarra:2001ju}
  F.~S.~Navarra, M.~Nielsen and M.~E.~Bracco,
  $D^* D\pi$ form-factor revisited,
  Phys.\ Rev.\ D {\bf 65}, 037502 (2002).

\bibitem{Aceti:2014uea}
  F.~Aceti, M.~Bayar, E.~Oset, A.~Martinez Torres, K.~P.~Khemchandani, J.~M.~Dias, F.~S.~Navarra and M.~Nielsen,
  Prediction of an $I=1$ $D \bar D^*$ state and relationship to the claimed $Z_c(3900)$, $Z_c(3885)$,
  Phys.\ Rev.\ D {\bf 90}, 016003 (2014).


\bibitem{Chen:2010ze}
  W.~Chen and S.~L.~Zhu,
  The Vector and Axial-Vector Charmonium-like States,
  Phys.\ Rev.\ D {\bf 83}, 034010 (2011).

\bibitem{Zhao:2014qva}
  L.~Zhao, W.~Z.~Deng and S.~L.~Zhu,
  Hidden-Charm Tetraquarks and Charged $Z_c$ States,
  Phys.\ Rev.\ D {\bf 90}, 094031 (2014).

\bibitem{Voloshin:2013dpa}
  M.~B.~Voloshin,
  $Z_c(3900)$ - what is inside?,
  Phys.\ Rev.\ D {\bf 87}, 091501 (2013).

\bibitem{Patel:2014vua}
  S.~Patel, M.~Shah and P.~C.~Vinodkumar,
  Mass spectra of four-quark states in the hidden charm sector,
  Eur.\ Phys.\ J.\ A {\bf 50}, 131 (2014).

\bibitem{Faccini:2013lda}
  L.~Maiani, V.~Riquer, R.~Faccini, F.~Piccinini, A.~Pilloni and A.~D.~Polosa,
  A $J^{PG}=1^{++}$ Charged Resonance in the $Y(4260) \to \pi^+ \pi^- J/\psi$ Decay?,
  Phys.\ Rev.\ D {\bf 87}, 111102 (2013).

\bibitem{Deng:2015lca}
  C.~Deng, J.~Ping, H.~Huang and F.~Wang,
  Systematic study of Z$_c^+$ family from a multiquark color flux-tube model,
  Phys.\ Rev.\ D {\bf 92}, 034027 (2015)






\bibitem{Szczepaniak:2015eza}
  A.~P.~Szczepaniak,
  Triangle Singularities and XYZ Quarkonium Peaks,
  Phys.\ Lett.\ B {\bf 747}, 410 (2015).

\bibitem{Chen:2011xk}
  D.~Y.~Chen and X.~Liu,
  Predicted charged charmonium-like structures in the hidden-charm dipion decay of higher charmonia,
  Phys.\ Rev.\ D {\bf 84}, 034032 (2011).

\bibitem{Chen:2013coa}
  D.~Y.~Chen, X.~Liu and T.~Matsuki,
  Reproducing the $Z_c(3900)$ structure through the initial-single-pion-emission mechanism,
  Phys.\ Rev.\ D {\bf 88}, 036008 (2013).

\bibitem{Swanson:2014tra}
  E.~S.~Swanson,
  $Z_b$ and $Z_c$ Exotic States as Coupled Channel Cusps,
  Phys.\ Rev.\ D {\bf 91}, 034009 (2015).

\bibitem{Swanson:2015bsa}
  E.~S.~Swanson,
  Cusps and Exotic Charmonia,
  Int.\ J.\ Mod.\ Phys.\ E {\bf 25}, 1642010 (2016).

\bibitem{Albaladejo:2015lob}
  M.~Albaladejo, F.~K.~Guo, C.~Hidalgo-Duque and J.~Nieves,
  $Z_c(3900)$: What has been really seen?,
  Phys.\ Lett.\ B {\bf 755}, 337 (2016).




\bibitem{Ablikim:2019ipd}
  M.~Ablikim {\it et al.} [BESIII Collaboration],
  Evidence for $Z_{c}^{\pm}$ decays into the $\rho^{\pm} \eta_{c}$ final state,
  arXiv:1906.00831 [hep-ex].


\bibitem{Goerke:2016hxf}
  F.~Goerke, T.~Gutsche, M.~A.~Ivanov, J.~G.~Korner, V.~E.~Lyubovitskij and P.~Santorelli,
  Four-quark structure of $Z_c(3900)$, $Z(4430)$ and $X_b(5568)$ states,
  Phys.\ Rev.\ D {\bf 94}, 094017 (2016).

\bibitem{Patel:2014zja}
  S.~Patel, M.~Shah, K.~Thakkar and P.~C.~Vinodkumar,
  Decay widths of Di-mesonic molecular states as candidates for $Z_{c}$ and $Z_b$,
  PoS Hadron {\bf 2013}, 189 (2013).

\bibitem{Esposito:2014hsa}
  A.~Esposito, A.~L.~Guerrieri and A.~Pilloni,
  Probing the nature of $Z_c^{(')}$ states via the $\eta_c \rho$ decay,
  Phys.\ Lett.\ B {\bf 746}, 194 (2015).

\bibitem{Ke:2013gia}
  H.~W.~Ke, Z.~T.~Wei and X.~Q.~Li,
  Is $Z_c(3900)$ a molecular state,
  Eur.\ Phys.\ J.\ C {\bf 73},2561 (2013).




\bibitem{Dias:2013xfa}
  J.~M.~Dias, F.~S.~Navarra, M.~Nielsen and C.~M.~Zanetti,
  $Z^+_c$(3900) decay width in QCD sum rules,
  Phys.\ Rev.\ D {\bf 88}, 016004 (2013).

\bibitem{Wang:2017lot}
  Z.~G.~Wang and J.~X.~Zhang,
  The decay width of the $Z_c(3900)$ as an axialvector tetraquark state in solid quark-hadron duality,
  Eur.\ Phys.\ J.\ C {\bf 78}, 14 (2018).

\bibitem{Agaev:2016dev}
  S.~S.~Agaev, K.~Azizi and H.~Sundu,
  Strong $Z_c^{+}(3900)\rightarrow J/\psi \pi^{+}; \eta_{c} \rho^{+}$ decays in QCD,
  Phys.\ Rev.\ D {\bf 93}, 074002 (2016).

\bibitem{Chen:2019wjd}
  H.~X.~Chen,
  Decay properties of the $Z_c(3900)$ through the Fierz rearrangement,
  arXiv:1910.03269 [hep-ph].











%
%
%
%
%
%
%
%
%
%
%
%
%
%
%
%
%
%
%
%
%
%
%
%
%

%
%
%
%
%
%
%
%
%





\bibitem{Barnes:2000hu}
  T.~Barnes, N.~Black and E.~S.~Swanson,
  Meson meson scattering in the quark model: Spin dependence and exotic channels,
  Phys.\ Rev.\ C {\bf 63}, 025204 (2001).

\bibitem{Barnes:1999hs}
  T.~Barnes, N.~Black, D.~J.~Dean and E.~S.~Swanson,
  $B B$ intermeson potentials in the quark model,
  Phys.\ Rev.\ C {\bf 60}, 045202 (1999).

\bibitem{Hilbert:2007hc}
  J.~P.~Hilbert, N.~Black, T.~Barnes and E.~S.~Swanson,
  Charmonium-Nucleon Dissociation Cross Sections in the Quark Model,
  Phys.\ Rev.\ C {\bf 75}, 064907 (2007).

\bibitem{Swanson:1992ec}
  E.~S.~Swanson,
  Intermeson potentials from the constituent quark model,
  Annals Phys.\  {\bf 220}, 73 (1992).

\bibitem{Barnes:1991em}
  T.~Barnes and E.~S.~Swanson,
  A Diagrammatic approach to meson meson scattering in the nonrelativistic quark potential model,
  Phys.\ Rev.\ D {\bf 46}, 131 (1992).

\bibitem{Wong:2001td}
  C.~Y.~Wong, E.~S.~Swanson and T.~Barnes,
  Heavy quarkonium dissociation cross-sections in relativistic heavy ion collisions,
  Phys.\ Rev.\ C {\bf 65}, 014903 (2002).

\bibitem{Godfrey:1985xj}
  S.~Godfrey and N.~Isgur,
  Mesons in a Relativized Quark Model with Chromodynamics,
  Phys.\ Rev.\ D {\bf 32}, 189 (1985).


%



\bibitem{Weinberg:1962hj}
  S.~Weinberg,
  Elementary particle theory of composite particles,
  Phys.\ Rev.\  {\bf 130}, 776 (1963).

\bibitem{Weinberg:1963zza}
  S.~Weinberg,
  Quasiparticles and the Born Series,
  Phys.\ Rev.\  {\bf 131}, 440 (1963).




\end{thebibliography}
\end{document}